\documentstyle[prl,aps,twocolumn]{revtex}
\input{epsf}
\begin{document}

\title{Nesting properties and anisotropy of the Fermi surface of
LuNi$_{2}$B$_{2}$C}

\author
{S.B. Dugdale, M.A. Alam, I. Wilkinson and R.J. Hughes}

\address{H.H. Wills Physics Laboratory, University of Bristol, Tyndall 
Avenue, Bristol BS8 1TL, United Kingdom}

\author{I.R. Fisher and P.C. Canfield}

\address{Ames Laboratory and Department of Physics and Astronomy, Iowa
State University, Ames, Iowa 50011}

\author{T. Jarlborg and G. Santi}

\address{D\'epartement de Physique de la Mati\`ere Condens\'ee,
Universit\'e de Gen\`eve, 24 quai Ernest Ansermet, CH-1211 Gen\`eve 4,
Switzerland} 

\date{\today}
\maketitle
\begin{abstract}

The rare earth nickel borocarbides, with the generic formula
$R$Ni$_{2}$B$_{2}$C, have recently been shown to display a rich variety of
phenomena. Most striking has been the competition between, and even
coexistence of, antiferromagnetism and superconductivity. We have measured
the Fermi surface (FS) of LuNi$_{2}$B$_{2}$C, and shown that it possesses
nesting features capable of explaining some of the phenomena experimentally
observed. In particular, it had previously been conjectured that a
particular sheet of FS is responsible for the modulated magnetic structures
manifest in some of the series. We report the first direct experimental
observation of this sheet.

\end{abstract}


The discovery of superconductivity in the rare earth nickel borocarbides
\cite{nagarajan:94} has raised fundamental questions regarding its nature.
The presence of moment-bearing rare-earth atoms in addition to high Ni
contents makes the very existence of superconductivity surprising, and some
of the phenomena associated with its interplay with magnetism have not
been observed in any other superconducting material. With the generic
formula $R$Ni$_{2}$B$_{2}$C, some members of the system ($R=$Y, Lu, Tm, Er,
Ho and Dy), exhibit moderately high values of T$_{c}$ ($\sim$16K, for
$R$=Y,Lu), and are either antiferromagnetic or non-magnetic at low
temperature. Two other systems, with $R=$Gd, Tb, are not superconducting,
while the Yb compound shows heavy fermion behavior \cite{canfield:98}.

One of the most striking observations has been the antiferromagnetic
ordering, and its competition (and even coexistence) with
superconductivity, conjectured to be driven by a nesting feature in the
Fermi surface (FS) \cite{rhee:95}.  Neutron and x-ray scattering techniques
have revealed incommensurate magnetic structures in superconducting
Er\cite{zaretsky:95} and Ho\cite{goldman:94} compounds, and in the
non-superconducting Tb\cite{dervenagas:96} and Gd\cite{detlefs:96}
compounds, characterized by a wave vector ${\mathbf Q_{\mbox{m}}} \approx
(0.55,0,0)$. In the Lu compound, the 4$f$ band is fully occupied and
therefore it is non-magnetic.  Since the $f$ electrons occupy localized
core-like states, the FS is expected to be similar to that of the other
compounds \cite{rhee:95}; unfettered by the complications introduced by the
presence of magnetism, the Lu compound is ideal for investigating the
origin of the superconductivity in these materials. Where there is magnetic
order, the mutual influence of the moments must occur through indirect
Ruderman-Kittel-Kasuya-Yosida (RKKY) type interactions; the resulting
magnetic structures would then be determined by maxima in the generalized
magnetic susceptibility of the conduction electrons.  Band theoretical
calculations of this susceptibility, $\chi(\mathbf{q})$, in
LuNi$_{2}$B$_{2}$C show a peak around the aforementioned vector, ${\mathbf
Q_{\mbox{m}}}$, indicating that the magnetic ordering in those other
compounds is a result of a common FS nesting feature
\cite{rhee:95}. Moreover, the presence of strong Kohn anomalies in the
phonon dispersion curves of LuNi$_{2}$B$_{2}$C for wave vectors close to
${\mathbf Q_{\mbox{m}}}$ has lent additional support to this conjecture
\cite{dervenagas:95}. Interest in the FS topology has been enhanced by the
observation of a four-fold symmetry in the anisotropy of the upper critical
field of LuNi$_{2}$B$_{2}$C \cite{metlushko:97}. This has been interpreted
in terms of (a) the anisotropy of the underlying FS topology, and (b) the
presence of possible three-dimensional $d$-wave superconductivity
\cite{wang:98}. In the absence of any corroborative evidence for the latter
suggestion, we shall concentrate on the former. The appearance of the
square flux-line lattice (FLL) has been successfully explained through
nonlocal corrections to the London theory \cite{kogan:96}. In general,
there is a nonlocal relationship between the supercurrent, $\mathbf{j}$,
and the vector potential, $\mathbf{A}$, of the magnetic field in a
superconductor, arising from the spatial extent ($\sim \xi_{0}$) of the
Cooper pair \cite{tinkham:96}. In this scenario, the shape of the FS (which
influences the coupling of the supercurrents with the crystal lattice) is
responsible for the square FLL observed at high fields in the mixed state
and which reverts to the triangular FLL at lower fields \cite{yaron:96}.
However, despite extensive theoretical work based on the band structures of
the non-magnetic Y and Lu compounds
\cite{mattheiss:94,lee:94,pickett:94,rhee:95}, very few experimental
determinations of the electronic structure exist, and many of the
predictions, including a ``nestable'' FS sheet, remain unverified. Most
importantly, the de Haas-van Alphen experiments performed on
YNi$_{2}$B$_{2}$C in the superconducting \cite{goll:96} and normal states
\cite{nguyen:96} have not delivered the topological information necessary
to isolate any such features in the FS.

In this Letter, we present a joint experimental and theoretical study of
the FS of LuNi$_{2}$B$_{2}$C. The experiment reveals the first direct
evidence for the presence of a sheet capable of nesting
\cite{rhee:95}. Furthermore, we will show that the anisotropy of this FS,
as determined by our band structure calculations, is consistent with
Kogan's model for the FLL \cite{kogan:96}.

The occupied momentum states, and hence the FS, can be accessed via the
momentum distribution using the 2-Dimensional Angular Correlation of
electron--positron Annihilation Radiation (2D-ACAR) technique
\cite{west:95}. A 2D-ACAR measurement yields a 2D projection (integration
over one dimension) of an underlying two-photon momentum density,
\begin{eqnarray}
\rho({\mathbf p}) &=& 
\sum_{j,{\mathbf k},{\mathbf G}} n^{j}({\mathbf k}) \vert
C_{{\mathbf G},j}({\mathbf k}) \vert ^{2}
\delta({\mathbf p}- {\mathbf k}-{\mathbf G}),
\end{eqnarray}
where $n^{j}({\mathbf k})$ is the electron occupation density in ${\mathbf
k}$-space in the $j^{\mbox{th}}$ band, the $C_{{\mathbf G},j}({\mathbf k})$
are the Fourier coefficients of the interacting electron-positron wave
function product and the delta function expresses the conservation of
crystal momentum.  $\rho({\mathbf p})$ is a single-centered distribution
having the full point symmetry of the the crystal lattice in question. In a
metal, this distribution contains discontinuities at various points
${\mathbf p}_{\mbox{F}} = ({\mathbf k}_{\mbox{F}} + {\mathbf G})$ when a
band crosses the Fermi level, $E_{\mbox{F}}$.  When the FS is 
of paramount interest, the Lock-Crisp-West procedure \cite{lock:73} is
often followed. Here the various FS discontinuities are superimposed by
folding $\rho({\mathbf p})$ (or its measured projections) back into the
first Brillouin zone (BZ). The result is a new ${\mathbf k}$-space density,
$\sum_{j,{\mathbf k}} n^{j}({\mathbf k})\sum_{{\mathbf G}} \vert
C_{{\mathbf G},j}({\mathbf k}) \vert ^{2}$, which aside from the factor
$\sum_{G} \vert C_{{\mathbf G},j}({\mathbf k}) \vert ^{2}$ (usually a weak
function of ${\mathbf k}$ within each band) is simply the electron
occupation density.  This well-established technique has recently been used
by some of the present authors to identify and measure the FS topology in
pure Y \cite{dugdale:97} and in disordered Gd--Y alloys
\cite{fretwell:99}. The virtue of the 2D-ACAR technique in such
studies is that it reveals {\it directly} the shape of the FS, and hence
any propensity for nesting.

The experiments were performed on a single crystal of LuNi$_{2}$B$_{2}$C,
grown by a Ni$_{2}$B flux method \cite{xu:94}. The sample was cooled to
$\sim$50K, at which temperature the overall momentum resolution of the
Bristol 2D-ACAR spectrometer corresponded to $\sim$10\% of the larger BZ
dimension.  Projections were measured along two different crystallographic
directions, $a$ and $c$. More than 400 million (effective) counts were
collected in each spectrum.

The spin-dependent momentum densities were calculated using the linearized
muffin-tin orbital (LMTO) method within the atomic sphere approximation
(ASA), including combined-correction terms \cite{andersen:75}. The
exchange-correlation part of the potential was described in the local
density approximation. The self-consistent band-structure was calculated at
594 $k$-points in the irreducible 1/16$^{th}$ part of the BZ, which to
simplify the calculations had the same volume as the standard BZ but a
simpler tetragonal shape. We used a basis set of $s$, $p$, $d$ and $f$
functions for the Lu, and a reduced ($s$,$p$,$d$) basis set for Ni, B and
C. For the calculation of the density-of-states (DOS), and the FS, a denser
mesh of 3696 $k$-points was used.  The lattice parameters used were
experimental values of $a=3.464$~\AA~ and $c=10.631$~\AA~
\cite{siegrist:94}. Electronic wave functions from the dense $k$-mesh
were then used to generate the electron-positron momentum densities, using
1149 reciprocal lattice vectors. A full description of the technique is
given in Refs.~\cite{andersen:75} and \cite{singh:85}.

\begin{figure}
\epsfxsize=180pt
\epsffile{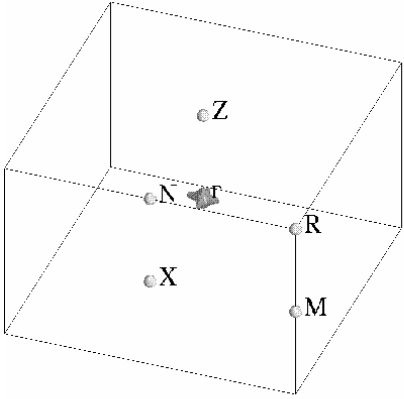}
\epsfxsize=180pt
\epsffile{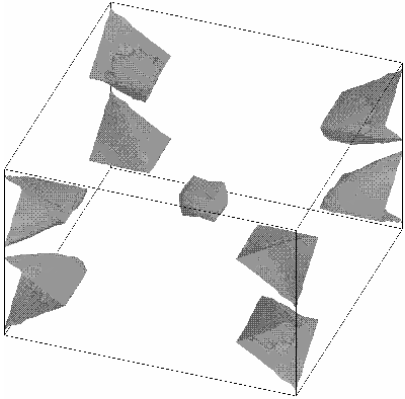}
\epsfxsize=180pt
\epsffile{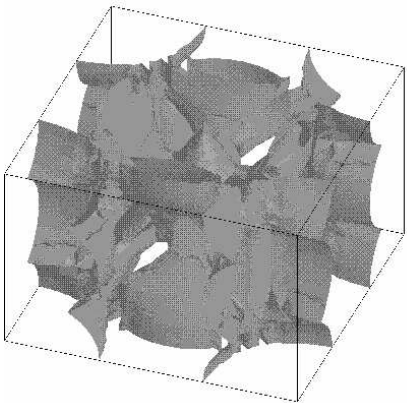}
\caption{The three sheets of the FS of LuNi$_{2}$B$_{2}$C, shown within the
simple tetragonal BZ (see text), with its usual symmetry points
labelled.}
\label{fs}
\end{figure}

The band-structure obtained (not shown) is very similar to those of Pickett
and Singh \cite{pickett:94} and Mattheiss \cite{mattheiss:94}, and shows
that the electronic structure is certainly three-dimensional (compared to
the two-dimensional features exhibited in the high-T$_{c}$ cuprates,
e.g. \cite{ito:91}); experimentally, one observes an almost isotropic
resistivity \cite{fisher:97}, which is consistent with this picture.  These
bands predict a rather complicated FS (shown in Fig.~\ref{fs}), the
principal character being Lu $d$, with some Ni $d$, and B and C $p$
character \cite{mattheiss:94,pickett:94}.  The first sheet is a very small
electron pocket centered at $\Gamma$, while the second is slighly larger
and rather more cuboid in shape. It is the third sheet that possesses the
nesting properties previously remarked upon \cite{rhee:95}, but it can also
be seen that the second sheet has a square cross section between $M$ and
$R$.

\begin{figure}
\epsfxsize=230pt
\epsffile{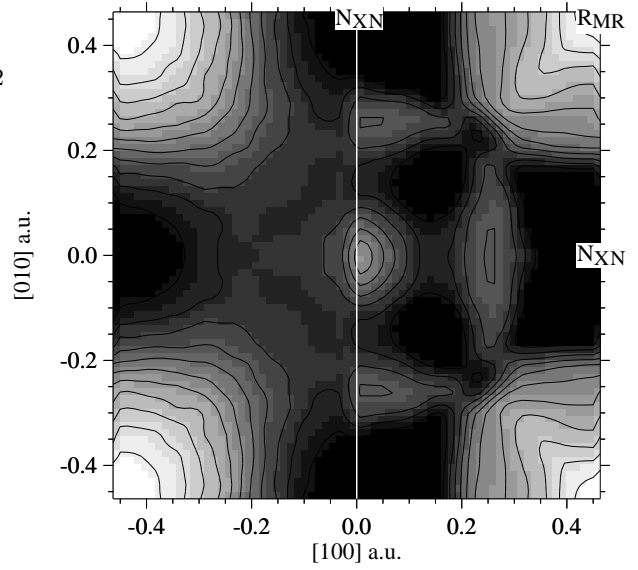}
\caption{Experimental (left) and calculated electron density (right) projected
along the [001] direction, with the symmetry points of the simple
tetragonal BZ indicating the projection path. Black signifies holes, and
white represents electrons.}
\label{cproj}
\end{figure}

\begin{figure}
\epsfxsize=230pt
\epsffile{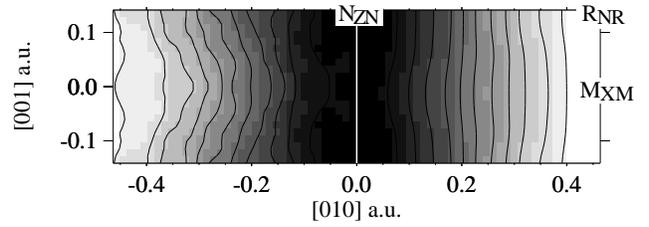}
\caption{Experimental (left) and calculated electron density (right) projected
along the [010] direction.}
\label{aproj}
\end{figure}

In Figs.~\ref{cproj} and \ref{aproj}, the BZ electron occupancies (both
calculated and experimental) are shown, projected along the $c$ and $a$
axes, respectively. The occupied states are indicated by the white areas,
and unoccupied by black. The agreement between the positron experiment and
the LMTO theory is excellent; one can clearly discern the presence of the
$\Gamma$-centered electron sheets. The most striking features of
Fig.~\ref{cproj} are the electron surfaces with square cross-sections at
the corners of the projected BZ. Also noteworthy are the shapes described
by the contour lines around this sheet; the protuberances point
towards the $\Gamma$-point in both the calculation and
experiment. Fig.~\ref{aproj} shows fewer features, but demonstrates that
the agreement persists when the projection is along a different direction.

\begin{figure}
\epsfxsize=230pt
\epsffile{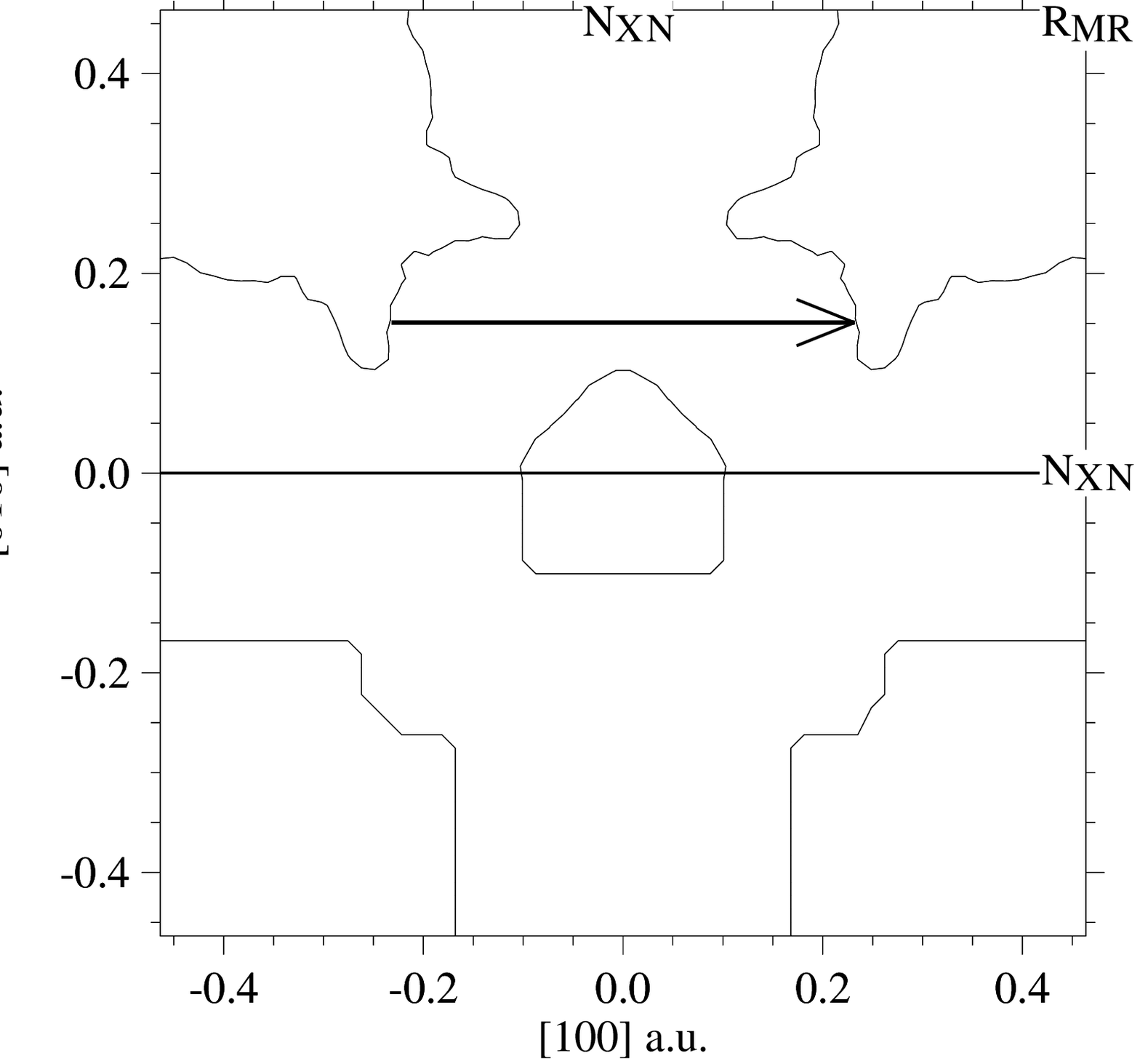}
\caption{The experimental (top) and calculated (bottom) FS topology of
LuNi$_{2}$B$_{2}$C. The calculation is of the FS in the third band in the
(001) plane through the $\Gamma$-point. The arrow indicates the nesting
feature.}
\label{cprojfs}
\end{figure}

Considerable investment has been made in establishing
methods of reliably and accurately locating the FS in 2D-ACAR data
\cite{dugdale:94}. Recently, it has been demonstrated that it is possible
to accurately caliper the FS from 2D-ACAR data using a criterion based upon
edge-detection algorithms employing band-pass filtering or Maximum Entropy
techniques \cite{dugdale:97,fretwell:99,livesay:99}.  In Fig.~\ref{cprojfs}
we present the experimental FS thus derived \cite{dugdale:97,fretwell:99},
together with a (001) section (passing through the $\Gamma$-point) of the
calculated third FS sheet. The nesting is indicated by the arrow, and is
calipered from the experimental data to be $0.54 \pm 0.02 \times (2\pi/a)$
(our calculation gives $0.56 \times (2\pi/a)$). Given that the
experimental data represent a {\it projection} of the FS, one does
not expect a perfect match between the top and bottom halves of the
figure. However, the nesting feature is unequivocally
revealed, and its size is in excellent agreement with the pronounced Kohn
anomalies close to a wave vector ${\mathbf Q_{\mbox{m}}}\approx (0.5,0,0)$
\cite{dervenagas:95}, and the observed incommensurate ordering with
${\mathbf Q_{\mbox{m}}}\approx (0.55,0,0)$ found in the antiferromagnetic
compounds. In conjunction with the calculation, it was possible to estimate
from the experiment that the fraction of the FS that would be able to
participate in nesting is $(4.4 \pm 0.5)$ \%, which is consistent
with the small increase in the resistivity observed when the current is
along [100] \cite{budko:99}. 

Having shown that the calculated FS topology is in good agreement with our
experiment, we derived pertinent quantities from our band structure. The
DOS at the Fermi energy, $N(E_{F})=4.3$ (eV cell)$^{-1}$, is similar to
values presented by Pickett and Singh \cite{pickett:94} and Mattheiss
\cite{mattheiss:94} (4.8 (eV cell)$^{-1}$).  Most importantly, some of the
quantities essential to Kogan's theory \cite{kogan:96} can be derived from
the band structure itself. For the calculation of the Fermi velocities, a
special mesh was used, comprising six additional points around each
original $k$-point, to enable accurate evaluation of the relevant
derivatives.  The Fermi velocities in the $ab$ plane and in the $c$
direction \cite{fulllist} were $\langle v_{F,ab}^{2} \rangle ^{1/2} = 2.62
\times 10^{7}$cm s$^{-1}$,$\langle v_{F,c} ^{2} \rangle ^{1/2} = 2.23
\times 10^{7}$cm s$^{-1}$, with masses $m_{ab}=0.91$, $m_{c}=1.26$,
implying an average out-of-plane anisotropy in the upper critical field
($(H^{<100>}_{c2} + H^{<110>}_{c2})/2H^{<001>}_{c2}$) of 1.17 compared to
an experimental value of 1.16 \cite{metlushko:97}. A calculation for the
isovalent compound YNi$_{2}$B$_{2}$C predicted 1.02 for this anisotropy, in
good agreement with the experimental value of 1.01 \cite{johnston:95} and
highlighting its sensitivity to the band structure.

In conclusion, we have shown experimentally that the FS topology of
LuNi$_{2}$B$_{2}$C does support nesting, thereby accounting for the
anomalies observed in its phonon spectrum \cite{dervenagas:95}, and the
propensity for magnetic ordering found in some of the other rare-earth
nickel borocarbides \cite{canfield:98}. In addition, our own calculations
of the electronic structure yield a FS in excellent agreement with the
experiments, whose anisotropy is consistent with the observation of a
square FLL, and with the observed behavior of the upper critical field
\cite{metlushko:97}.

The authors would like the thank the EPSRC (UK) for financial support, and
B. Harmon and R. Evans for useful discussions. One of us (SBD) 
ackowledges support from the Lloyd's of London Tercentenary
Foundation. Ames Laboratory is operated for the U.S. Department of Energy
by Iowa State University under Contract No. W-7405-Eng-82.  This work was
supported by the Director for Energy Research, Office of Basic Energy
Sciences.

\begin{thebibliography}{99}
\bibitem{fulllist}
The Fermi velocity averages are as follows~: 
$\langle v^{2} \rangle = 1.88 \times 10^{15}$ (cm/sec)$^{2}$,
$\langle v_{x}^{4} \rangle = 2.21 \times 10^{30}$ (cm/sec)$^{4}$,
$\langle v_{x}^{2}v_{y}^{2} \rangle = 1.63 \times 10^{29}$ (cm/sec)$^{4}$,
$\langle v_{z}^{4} \rangle = 6.97 \times 10^{29}$ (cm/sec)$^{4}$,
$\langle v_{x}^{2}v_{z}^{2} \rangle = 5.46 \times 10^{29}$ (cm/sec)$^{4}$.
\end{thebibliography}
\end{document}